\documentclass[pre,preprint,aps,]{revtex4}
\usepackage{subcaption}
\usepackage{graphicx}
\usepackage{comment}
\usepackage{amsmath,amssymb,amsthm} 
\usepackage{bm}
\usepackage{verbatim}
\usepackage{float}

\begin{document}
\title{Origin of homochirality: The formation and stability  of homochiral peptides in aqueous prebiological environment in the Earth's crust}
\author{ S\o ren  Toxvaerd  }
\affiliation{ Department
 of Science and Environment, Roskilde University, Postbox 260, DK-4000 Roskilde, Denmark, st@ruc.dk}
\date{\today}

\vspace*{0.7cm}

\begin{abstract}
The oldest forms of living organisms on Earth are  about 3,5 billion years old, and they are found  in hydrothermal  deposits,
and it is often hypothesized   that life originated
there. But the hydrothermal systems with a fairly strong flow  of  chemical components are  not the optimal place for the prebiological  self-assembly of
biomolecules and for the emergence of homochirality. This article examines the possibility for that the self-assembly
of homochiral molecules  took place in an aqueous environment  in the Earth's crust.
Based on the latest literature regarding the conditions
in the lithosphere there  are several factors that point to that the crust could be the
location for the prebiological self-assembly of biomolecules, and there is nothing against it. The crust and the mantle contain a substantial amount of water,
	and  at the time prior to
the emergence of life the crust contained most likely the
necessary chemical substances for the synthesis of the biomolecules and an aqueous environments where the homochirality could be established.
\end{abstract}

\maketitle
\vspace{2pc}

\section{Introduction}

Theories about the origin of life often deal
 with the emergence
of homochiral peptides of L-amino acids and homochiral oligomers of D-carbohydrates.
But not only are the chiral amino acids unstable  and with a tendency to racemize \cite{Bada1972},
it is  the peptides too  \cite{Aki2013,Fujii2018}. The homochirality in the proteins is vital
for the survival of an organism, and the racemization of Aspartic acid in endospores \cite{Liang2019} and hyperthermophiles  \cite{Onstott2014,Liang2021}  limits
the long-time survivability of  microorganisms.
So the fundamental
question is not only how to get the homochiral peptides, but also how to maintain the homochirality in the peptides
over so   long  a time that a self-assembly of the complex homochiral molecules  in  living systems, can
take place, and a living organism can be created.

The racemization of amino acids ranges from a time span of thousand  years at low temperatures
to days at 100 $^\circ C$, and the racemization  is enhanced in acidic as well as basic solutions \cite{Bada1985}.
Even if there somewhere was a pool of water with pH $\approx$ 7 of homochiral amino acid,
a simple  synthesis of homochiral peptides from an aqueous solution of homochiral amino acids
can only happen, if the synthesis of biomolecules with homochiral amino acids was completed
within a few thousand years in cold water or within some days in a hot aqueous solution.
But since no one can imagine that it took so short a time
to establish the prebiolocical environment with synthesis of the complex biomolecules,  one must search for  environments with the right conditions, where
  \textit{it is possible to obtain homochiral peptides from a racemic mixture of amino acids and
maintain the environment for million of years}, sufficient to ensure the self-assembly of the  homochiral
  building blocks in the living organisms.
  
  This  article focuses
  on a possibility for such an environment with   formation and stability  of homochiral peptides   in an aqueous solution. 
 If it is possible to  synthesize homochiral  peptides $from$ $a$ $racemic$ $mixture$ $of$ $amino$ $acids$  in an inorganic aqueous solution
 under the $right$  $conditions$, this can explain
 the emergence of homochirality in peptides  \cite{Toxvaerd2017}.

\section{The right conditions}
Most theories about the emergence of  homochirality of the peptides  take their starting point from  synthesis of peptides from
homochiral amino acids. But this is as mentioned unrealistic, and furthermore it is not necessary.
Kinetic models indicate that one can obtain homochiral peptides from
spontaneous polymerization  of racemic mixtures of amino acids \cite{Brandenburg,Ribo2017,Blanco2017,Buhse2022,Pineros2022}.
The conditions for obtaining homochiral peptides  from racemic compositions can be specified. There are  the $thermodynamic$ $conditions$ \cite{Toxvaerd2009}, but there are
also other conditions, e.g. to \textit{the chemistry at the Abiogenesis}, and to the \textit{location} with the right thermodynamic- and chemical environment.

\subsection{The right location}

All living systems consist of cells with an aqueous solution (cytosol) of ions and  suspensions of biomolecules, and this
simple fact makes some demands on the environment and the location where the self-assembly of the biomolecules appeared \cite{Toxvaerd2019}.
One  demand to the location of the environment is that it must have been an aqueous environment, and that  life must have arose in water.
Today the water on Earth is present  in the oceans, in the lakes, and in many other locations in  our biosphere, but  the main part of water on Earth is 
present in the crust and  mantle  beneath the continents and the oceans  \cite{Ohtani2021}.

Earth's lithosphere constitutes the hard and rigid outer  layer of the Earth and includes the crust and the uppermost mantle.
The crust is the   outermost solid  layer of the Earth below the oceans and the continents.  The  estimates of thickness of the crust vary from $\approx$  10 km
below the oceans to $\approx$ 70 km below the continents. The crust is separated from the upper part of the mantle by the ''Moho discontinuity" by a distinct
change of the density. The temperature in the lithosphere increases  with the distances from the Earth's surface,
and reaches a temperature of more than thousand degree at $\approx$  200 km \cite{Jeanloz1986}. The synthesis and the stability of peptides requires,
however, a significant lower temperature
\cite{Vogt1996,Shao2010}.
The  crust, mantle and the ''Hadean ocean" were presumably formed  relative shortly after the
 Earth was created $\approx$ 4.56 billion years ago  \cite{Miyazaki2022},
 and the moon, that was formed shortly after the formation of the Earth, was much closer than it is today 
  and with very strong tide waves in the Hadean ocean \cite{Barboni2017,Thiemens2019,Green2017} . (The strong tide waves 
   have  accelerated the Moon out to its present position.)
Today the water on Earth is present in 
 the oceans, in the lakes, and in many other locations in  our biosphere, but  the main part of water on Earth is 
present in the crust and  mantle  beneath the continents and the oceans. Part of the water
in the crust and the mantle is bound as crystal water, $\textrm{H}_2\textrm{O}(c)$, but an essential part  is present as water $\textrm{H}_2\textrm{O}(aq)$.  
The  evidence   for water in the crust and mantle  is indirect and based on extensive
seismic velocity measurements, electrical and thermal conductivity measurements, geodetic data, mineral physics, geochemical data, and modeling \cite{Ohtani2021,Harrison2020}.
But although the  indications of water in the crust and the mantle there is, however, no direct determinations of whether $\textrm{H}_2\textrm{O}(aq)$
appears as capillary or   bulk water. 
The review \cite{Ohtani2021}  gives an estimate of the total amount of water 
of 2.6-8.3 ocean  masses ( $\textrm{H}_2\textrm{O}$(c)+ $\textrm{H}_2\textrm{O}$(aq)).
The water in the crust and mantle is released in the hydrothermal vents, and the right place for prebiological self-assembly of organic molecules
could very well be the   $\textrm{H}_2\textrm{O}(aq)$
somewhere in the  Earth's crust.

A support for this hypothesis  is the fact that
the oldest signs of life, which are at least 3.5 billion years  old,  appear as  fossilized microorganisms (bacteria)
found in hydrothermal vent precipitates \cite{Djokic2017,Schopf2018,Cavalazzi2021,Dodd2017}.
There are also direct measurements of a biosphere in the upper part of the wet crust, which today contains bacterias \cite{ Li2020,Takamiya2021}.
Here we argue for that the the  location for prebiological self-assembly of homochiral peptides is a wet crust in the Hadean Eon. 

A  relationship that supports the hypothesis is the chemical composition of the cytosol.
 The cytosol contains sodium and potassium ions, and the  concentrations of potassium and 
 sodium in living systems
  differs significantly from the corresponding composition in the oceans.
 The ion content in
 the cytosol  points to a location in the  crust for  self-assembly of biomolecules
 and not to the Hadean ocean with a composition  of salt, which probably have been similar
 to the composition in today's oceans \cite{Harrison2009,Harrison2020a}.
  The  concentration of potassium  $[\textrm{K}_{\textrm{cytosol}}^+]\approx $0.10 M,  and the
  ratio between the concentration of potassium ions  and sodium ions\\
  $[\textrm{K}_{\textrm{cytosol}}^+]/[\textrm{Na}_{\textrm{cytosol}}^+]\approx 10 $ \\
  in the cytosol in the biocells deviate
  very much from the corresponding concentration of potassium  $[\textrm{K}_{\textrm{oceans}}^+]\approx $0.01022 M  and the ratio\\
   $[\textrm{K}_{\textrm{oceans}}^+]/[\textrm{Na}_{\textrm{oceans}}^+]=0.0102/0.469=0.0217  $\\
   in today's oceans.
   The ratio  $[\textrm{K}_{\textrm{oceans}}^+]/[\textrm{Na}_{\textrm{oceans}}^+]$ is crucial for the membrane potential and the sodium-potassium pump in the cells, and
   a high concentration of potassium and the right ratio between sodium and potassium
   could be present somewhere  in the crust and in mica sheets in the  upper part of the mantle \cite{Hansma2021}. Another
  example in this context is the appearance of phosphor esters in biological systems, in the membranes, in RNA, in DNA, in the Glycolysis, and in many
  other biomolecules. But whereas the concentration   of phosphates in the oceans in general are low 
  \cite{Paytan2007},  phosphates  is more common in the crust  \cite{Walton2021,Flores2022}.

   Capillaries are  locations where it, opposed to the oceans  is possible to obtain and maintain a 
  relative  high concentration of  amino acids.
  This is an important condition  for the  thermodynamics and kinetics at the prebiological bio-synthesis (see later).
  Many scientists have  suggested that life originated in the hydrodynamic
  vents and serpentinites  in the oceans \cite{Cleaves2009,Branscomb2018} in accordance with earliest sign of life on Earth.
  The environment in the hydrothermal vents and serpentinites  above the wet crust and mantle    is turbulent and
   with a rapid exchange of matter \cite{Moore2021}. If  the prebiological synthesis of the biomolecules appeared over a long  period of time it seems more likely,
  that a confined water system in the crust  with a small, but constant input of the chemical reactants in the biosynthesis, and with
  a slow synthesis of the biomolecules over a time period of many million of years, is the right location for prebiological synthesis of the biomolecules .
  
   This article examines the possibility, that the self-assembly of homochiral molecules  originated in an aqueous environment  in the Earth's crust.

 \subsection{The thermodynamic condition}

  A suspension of homochiral  peptides in an aqueous environment  is stable  if the peptides are 
 in structures  with minimum free energy.   The peptides are  polymers of  units of amino acids, combined via amide bonds. 
 The stability of peptides increases with temperature relative to hydrolysis reactions in aqueous solutions,
 and  become a facile process in hydrothermal systems  deep in sedimentary basins \cite{Shock1992,Takahagi2019,Pedreira-Segade2019}.
 A homochiral L-peptide in an aqueous ionic suspension is  stable if there is a sufficient ''chiral discrimination".
 The chiral discrimination for a homochiral L-peptide is given by the difference in  Gibbs free energies of the peptide and the peptide with
 one (or more) D- configuration  of its chiral units. This requirement can be specified:

 \textit{The increase in reaction enthalpy $ \Delta H_r(aq)$ by  a change of chirality  of one L-configuration
 to a D-configuration in the homochiral peptide in an aqueous ionic suspension shall exceed the decrease of reaction   energy  $-T \Delta S_r(aq)$
  by the gain in entropy.}

  The  excess of of reaction enthalpy   measured in units of $RT \approx $ 2.5 kJ/mol must be several units in order to ensure the stability
  of the homochiral conformation \cite{Toxvaerd2009}.
  As mentioned  especially the Aspartic acid units  in the proteins are  unstable and racemizate  with  consequence for the
  survivability  of biosystems  \cite{Liang2019,Liang2021}.

\begin{figure}
\begin{center}
\includegraphics[width=6cm,angle=0]{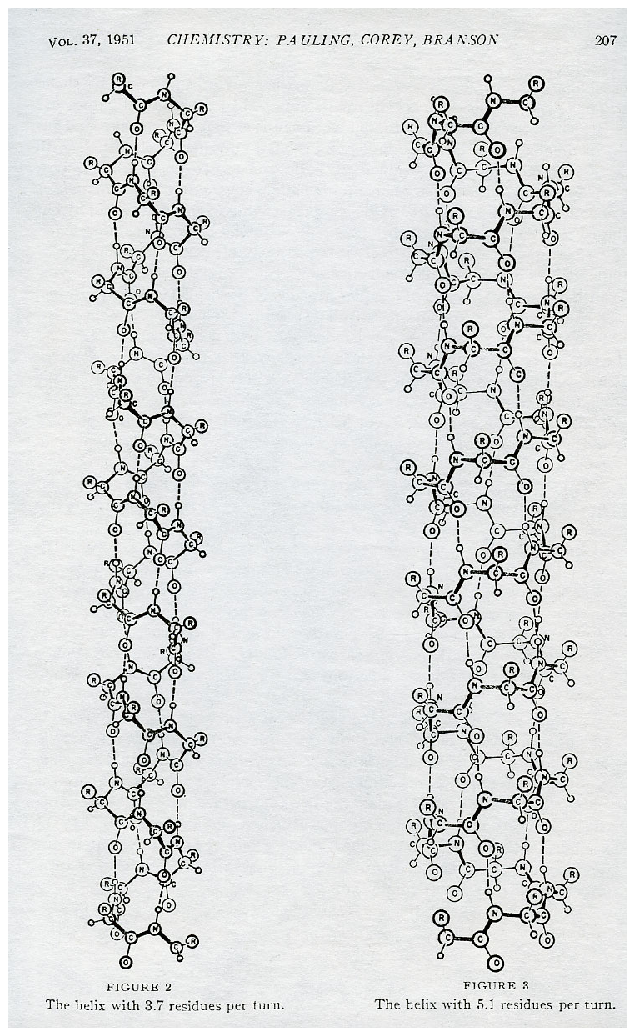}
\caption{The  secondary $\alpha$-helix  structure of a homochiral peptide chain \cite{Pauling1951}. The secondary structure is stabilized by
	(weak) hydrogen bonds. Each amide group is hydrogen bounded to the third (or fifth) amide group beyond
	it along the helix. The  $\alpha$-helix structure to the left of the figure is found in nature.}
\end{center}	
\end{figure}

The structure of a homochiral peptide is primarily given by its
 secondary structure, first determined by Pauling \cite{Pauling1951}. Perhaps the figure of the $\alpha$-helix
 structure in Pauling et al's article  best illustrates the condition for thermodynamic stability of a homochiral peptide in an aqueous suspension.
The  clockwise (positive) $\alpha$-helix structure (to the left) on Figure 1 with  amide units of L-amino acids was derived
from spectroscopic data for bond lengths and angles at the
planar peptide units, and with an $\it{intramolecular}$ stabilizing hydrogen-bond 
between a hydrogen atom at the substituted -N-H and an oxygen atom in a
-C=O group in the helix. The helix is   obtained for peptides with homochiral amino acids, by which the overall orientation changes with a constant 
angle at each peptide bond. The angle at the peptide bond  is changed  if an L configuration is changed to a D configuration, and the 
$\alpha$-helix structure is destroyed. This fact is confirmed  by the observation, that  a chiral change of an L-structure to a D-structure in one chiral unit
is associated with a loss of the secondary structure of the peptide and an increased concentration of water molecules in a peptide \cite{Fujii2010}.

The effect  of an ionic aqueous solution on the enthalpy  of a homochiral peptide  is not known, but one shall \textit{a priori} expect that the stability
of the $\alpha$-helix structure is stabilized by increasing ionic activity and thereby decreasing activity of the water molecules (Debye–H\"{u}ckel theory).
The energy of the -N-H$\cdot\cdot\cdot$O=C- hydrogen bond ($\approx$ 10 kJ/mol) is only
of the order of  one half of the energy of a
hydrogen bond in pure water \cite{Wendler2010}, and  the stabilizing intramolacular hydrogen bond is in competition with  hydrogen bonds to the water molecules in the  solvent. The ions in
the cytosol
decrease the  water activity and thereby the difference in energies  between the intramolecular  hydrogen bonds in the peptide and
a hydrogen bond with  water molecules in the cytosol.
 This fact is in agreement with the  simulations of simple chain molecules in a solvent with different activities \cite{Toxvaerd2017} which show, that
a decreasing activity of the solvent 
stabilizes the compact conformation of  the polymer.

There is another thermodynamic requirement to a prebiological  synthesis of peptides  in a wet crust.
  The  synthesis of the biomolecules in the aqueous environment requires on one hand a certain temperature, and on the other hand
  biomolecules are  unstable and decompose at high temperatures. 
  Only very little is, however, known about the temperature in the crust and mantle in the Hadean Eon \cite{Harrison2020a}.

 The thermodynamics requirement to the right place for maintaining peptides over long times in  homochiral secondary (or higher order) structures
 is an aqueous solution with a reduced water activity, given by the concentrations of the ions in the water. A high electrolyte concentration, higher than
  in the cytosol can be present in wet capillaries in the crust.

\subsection{The  conditions to the chemistry at the Abiogenesis}

The  conditions to the chemistry at the Abiogenesis  can be specified: There are the conditions
to the chemical reactions by which the homochiral peptides and 
carbohydrates are synthesized from racemic compositions, and there are the conditions to chemical composition  in the crust.
\subsubsection{The conditions to the chemical reactions: The Frank model}
The synthesis of homochiral  polymers can be obtained by consecutive reactions, first described by C. F. Frank  \cite{Frank1953}. 
In 1953  he published a theoretical model for spontaneous asymmetric synthesis, and the emergence of homochirality
at  the polymerization of peptides is an example of such a synthesis. Since then numerous of articles   
  with   asymmetrical polymerization and chiral amplification  have  been published  \cite{Brandenburg,Ribo2017,Blanco2017,Buhse2022,Pineros2022,Shen2021}.
An important  and necessary condition  to the network of consecutive reactions  is as mentioned the inclusion of
the racemization   of the monomers. This is included in a several of the reaction schemes in  the  articles \cite{Ribo2017,Blanco2017,Buhse2022,Pineros2022}, 
where the  kinetics reactions
for racemization of the chiral monomers are 
together with autocatalytic polymerization and merging of polymers, and the reaction equations with homochiral peptides
are solved for various values of reaction constants.

The asymmetric polymerization,  obtained by Frank models can lead to homo chiral peptides, but with equal possibility for
a both enantiomers (Pauling's polymers in Figure 1 are
in fact for secondary structures of  peptides with  D-amino acid units \cite{Dunitz2001}).
The  dominance of the L-conformations of peptides  may be caused  by
 the weak parity violating nuclear forces. This impact is, however, very small, and the dominance
 can also be obtained by a self-reinforcing dominance  of  homochiral domains.  \cite{Toxvaerd2000,Toxvaerd2009}. 

The  necessary chiral discrimination for a symmetry break and homochiral purification is indirectly given by  the ratios of the reaction constants \cite{Toxvaerd2000}.
  According to the thermodynamic condition in Section 2.2, an important details in the reaction scheme
for the consecutive polymerization reaction is the chiral discrimination, given by the hydrogen bonds in the secondary structure of the peptides in
an ionic aqueous solution.
The impact of these complicated effects of the conformation of the polymer in the ionic  environment is,
however, not   included in the Frank models for homochirality  obtained by  polymerization,
where the chiral discrimination from the reaction enthalpies is  given by the ratios of the reaction constants.

The  articles  with chiral amplification  are  for polymerization with uniform
(bulk) concentrations of the species   \cite{Brandenburg,Ribo2017,Blanco2017,Buhse2022,Pineros2022,Shen2021}.
An extension to  more realistic reaction-diffusion Turing models with local concentrations \cite{Turing1952} is, however,  complicated
and moreover  the homochiral purification might have been  obtained in  capillaries in the crust.
 A reaction-diffusion  Frank model  for the polymerization in a confined geometry 
is complicated and difficult to evaluate, but reaction-diffusion models for  consecutive  reactions in confined geometry have  
been developed and evaluated \cite{Hunding1990}. 

Chemical reactions in fluid phases are rather insensitive with respect to the pressure changes.
The  changes of partial Gibbs free energies with respect to
 a pressure change are given by the partial volumes of the components in the solution, and the Gibbs free energy effects by a pressure change  are   generally small.
The  polymerization to homochiral peptides might lead to a slightly
higher density of the suspension, an if so the increased pressure in the crust will enhance the polymerization rate, but it will not affect
the symmetry break because  
 homochiral suspensions with  peptides with  pure D- or L-units  have the same density.

In summary: The kinetics with racemization kinetics of the amino acids and with polymerization to  homochiral peptides
can be described by  a Frank model.
The thermodynamic stability of the homochiral peptides are ensured by intra-molecular  hydrogen bonds in the higher order structures in competition with
hydrogen bonds with the water molecules in the solvent, which destabilize the secondary structure. It is, however, difficult to take this effect into account in the kinetic
in the Frank models.

\subsubsection{The chemical composition}
The chemical composition of the Earth's  crust and upper part of the mantle  4-4.5 billion years ago is not known,
but today the solid mantle consists mainly of Mg, Si and Fe \cite{Anderson1983,Lyubetskaya2007} .
 In determining the chemical composition  of the crust and the mantle
 one utilizes cosmochemical data, based on the assumption that the sun, planets and meteorites all
originated from the solar nebular. Chondritic meteorites are considered the most representative samples
of the solid nebular material \cite{Lyubetskaya2007}. 
Today, the crust may only contains small amount of organic materials such as amino acids \cite{Pisapia2018} and carbohydrates  \cite{Sforna2018}, and 2.7 billion year old rocks
contained acetate and formate \cite{Lollar2021}. 

 The most interesting chondrites are the carbonaceous chondrites. Not only do they contain
 carbon, but also organic molecules an among these also amino acids and purine and pyrimidine nucleobases \cite{Oba2022}.
 There exists many investigations of the chemical composition in the
 carbonaceous chondrites and the composition of amino acids \cite{Aponte2020,Glavin2020}. The amino acids appear with a small excess of L-amino acids,
 which is to be expected since strong baryonic  forces in the universe favor the L-configurations.
 However, these small excesses of chirality are,  as mentioned irrelevant for the origin of  homochirality of peptides
 in aqueous solutions, because 
 the amino acids racemizate rapidly. But if a part of the crust or the mantle consisted of the same materials as the  carbonaceous chondrites some organic molecules
 might have survived the heating of the Earth at its formation, and the
 carbonaceous chondrites material might have been a source of amino acids at the polymerization of peptides in the crust.
 There are  also investigations 
 which makes it probable, that  amides can be directly synthesized  at thermodynamic conditions similar to the conditions in the crust \cite{Robinson2021,Fu2022,Yang2022}.

 Carbohydrates are  found in  carbonaceous chondrites \cite{Furukawa2019}, but they   can also be synthesized
 from carbon dioxide and methane, two component which
  also were present in the  nebular material  at the formation of the Earth and our solar system \cite{Segura2015}.
 The first step in the synthesis of the  carbohydrates  is formaldehyde, CH$_{2}$O,  which is synthesized
  from carbon dioxide and methane \cite{Schlesinger1983}, and the the succeeding  spontaneous  condensation of formaldehyde
  is the formose reaction,  which was discovered already in 1861 \cite{Butl}.
The formose reaction  is well known and is believed to be the basic synthesis of bio-carbohydrates
  and related bio-organic molecules  \cite{Gabel,Wash,Kim,Jalbout,Lambert}.
  The condensation into carbohydrates is catalyzed
 by not only amino acids \cite{Weber}; but also  by naturally
  aluminosilicates occurring in  the mantle \cite{Anderson1983}.

In summary: The  early lithosphere most likely contained  the basic chemical ingredients,
carbon dioxide and methane and ammonia, for synthesis of carbohydrates and peptides.

\section{Conclusion}
The oldest signs of cellular life on Earth are about 3,5 billion years old \cite{Djokic2017,Cavalazzi2021}, and they are found in a
volcanic-hydrothermal (hot springs, geysers) system  and in hydrothermal veins deposits, and it
 is often hypothesized   that life originated
there. But the hydrothemal systems with a fairly strong flow  of  chemical components are  not the optimal place for the prebiological  self-assembly of
biomolecules and for the emergence of homochirality. This article examines the possibility for that the self-assembly of homochiral molecules  originated in an aqueous environment  in the Earth's crust.

Based on the latest literature regarding the conditions
in the Earth's crust and upper part of the mantle there  are several factors that point to, that the crust could be the
location for the prebiological self-assembly of biomolecules, and there is nothing against it. 
 The crust and the upper part of the mantle  contains a substantial amount of water \cite{Ohtani2021}, and
 a substantial biomass and biodiversity \cite{Takamiya2021},  with
an estimated number of (Prokaryotic) cells of the order 2 to 6$\times 10^{29}$ \cite{Magnabosco2018}.
The crust and upper part of the mantle at the time prior to
the emergence of life contained  the
necessary chemical substances for the synthesis  of the biomolecules, and an aqueous environment
 in  the   crust  beneath hydrothermal locations  is a more likely place
for the prebiological synthesis of peptides and carbohydrates and for  the establishment of homochirality.

A wet crust can generally be a suitable place for the biochemical evolution primordial stages. Here we have advocated for the Earth's crust  in the
Hadean Eon
as the most likely place,  but it might as well have been on Mars. The requirement to a moderate temperature for synthesis of  stable biomolecules
sets a limit to how soon after the emergence of the solar system such a biosynthesis could take place, and this fact is in favour of Mars, which is a
terrestrial planet similar to Earth. It was created at the same time as the Earth, but  cooled down before the Earth \cite{Plesa2018}.

$\textbf{Acknowledgment}$

The anonymous, but constructive   referee reports are gratefully acknowledged.
This work was supported by the VILLUM Foundation’s Matter project, grant No. 16515.

$\textbf{References}$

\end{document}